\documentclass[manuscript]{aastex63}

\submitjournal{ApJ}

\shorttitle{Radiogenic Heating and Rocky Planet Dynamos}
\shortauthors{Nimmo et al.}
\graphicspath{{./}{figures/}}

\begin{document}

\title{ {
Radiogenic Heating and its Influence on Rocky Planet Dynamos and Habitability}}

\correspondingauthor{Francis Nimmo}
\email{fnimmo@ucsc.edu}

\author[0000-0000-0000-0000]{Francis Nimmo}
\affiliation{Department of Earth and Planetary Sciences, University of California Santa Cruz, California, USA}

\author{Joel Primack}
\affiliation{Physics Department, University of California Santa Cruz, 1156 High Street, 
Santa Cruz, CA 95064 USA}

\author{S. M. Faber}
\affiliation{Department of Astronomy, University of California Santa Cruz, 1156 High Street, 
Santa Cruz, CA 95064 USA}

\author{Enrico Ramirez-Ruiz}
\affiliation{Department of Astronomy, University of California Santa Cruz, 1156 High Street, 
Santa Cruz, CA 95064 USA; and Niels Bohr Institute, University of Copenhagen, Blegdamsvej 17, 2100 Copenhagen, Denmark}

\author{Mohammadtaher Safarzadeh}
\affiliation{Department of Astronomy, University of California Santa Cruz, 1156 High Street, 
Santa Cruz, CA 95064 USA}



\begin{abstract}
The thermal evolution of rocky planets \textcolor{black}{on geological timescales (Gyr)} depends on the heat input from the long-lived
radiogenic elements potassium, thorium, and uranium.  Concentrations of the latter two \textcolor{black}{in rocky planet mantles} are likely to vary by up to an order of magnitude between different planetary systems because Th and U, like other heavy r-process elements, are produced by rare stellar processes.  Here we discuss the effects of these variations on the thermal evolution of an Earth-size planet, using a \textcolor{black}{1D} parameterized convection model. Assuming Th and U abundances consistent with geochemical models of the Bulk Silicate Earth based on chondritic meteorites, we find that Earth had just enough radiogenic heating to maintain a persistent dynamo. \textcolor{black}{According to this model,} Earth-like planets of stars with \textcolor{black}{higher} abundances of heavy r-process elements, indicated by the relative abundance of europium in their spectra, are likely to have lacked a dynamo for a significant fraction of their lifetimes, with potentially negative consequences for \textcolor{black}{hosting a biosphere}. \textcolor{black}{Because the qualitative outcomes of our 1D model are
strongly dependent on the treatment of viscosity, further investigations using fully 3D convection
models are desirable.}

\end{abstract}



\section{Introduction} \label{sec:intro}

Rocky planets are gigantic heat engines \citep{Steve:2008}. They lose internal heat to the surface via some 
combination of conduction, convection and advection, driving internal processes such as dynamo generation, 
volcanism or plate tectonics as they do so. The internal heat arises in 
three ways: gravitational energy released as the planet formed; tidal heating, as for Io or some exoplanets \citep{HenniHurfo:2014}; and decay
of radioactive elements. Short-lived radionuclides such as $\rm ^{26}Al$ and $\rm ^{60}Fe$ may have contributed energy in the first \textcolor{black}{$\sim$3~Myr of our Solar System} and added to the primordial gravitational energy \citep[reviewed in][]{Lugaro:2018}. Long-lived radionuclides include $\rm ^{40}K$, $\rm^{232}Th$, $\rm ^{238}U$ and $\rm ^{235}U$ and are the focus of this work.  U and Th  have been found in the Milky Way to be pure r-process products with concentrations characteristic of Solar System matter but with sizable 
expected star-to-star bulk scatter in their concentrations with respect to lighter elements \textcolor{black}{such as Mg} \citep{2008ARA&A..46..241S,Cowan-etal:2019}.
As a result, we expect other planetary systems to posses quite different concentrations of U and Th. Some geological consequences of such radiogenic element variations have been 
explored hitherto \citep{Frank-etal:2014,Unter-etal:2015,JelliJacks:2015,FoleySmye:2018,Foley:2019,Botelho:2019,Quick-etal:2020,Wang-etal:2020}. \textcolor{black}{But to our knowledge the only paper examining how exoplanet dynamos would be affected by variations in U,Th is by \citet{ONeill-etal:2020}, who assume these variations are small and correlate with core size. For reasons explained below we assume much larger, stochastic variations unconnected to core size.}  

For an Earth-mass body, the gravitational energy released by accretion is roughly 40~MJ/kg, whereas the total energy 
released by long-lived radioactive decay from formation to the present is roughly 1~MJ/kg.  But while accretion energy is delivered in short intervals 
during the first few Myr and \textcolor{black}{thus sets the initial thermal conditions}, radiogenic heating is an ongoing process occurring mostly in the mantle and crust. As a result, it is the latter that
matters more for long-term geological evolution. 

Estimates of radiogenic heat production for planets in our Solar System rely heavily on the elemental concentrations
 found in primitive meteorites (see Appendix). For the Earth, samples of the crust and upper mantle and direct measurements using 
 geoneutrinos \citep{Agostini:2019} provide additional constraints. At the present day, plate tectonics causes the Earth
  to lose heat roughly twice as fast as it is being generated by radioactive decay \citep{Sclat-etal:1980}, so Earth is cooling. Four billion years
   ago, heat production was 3.5 times as large and the majority of heat transfer may have arisen through advection (volcanism) \citep{MooreWebb:2013}. 

In the rocky planets, magnetic fields are generated by convection in a metallic core, which in turn is driven by heat extracted into the
 mantle \citep{Nimmo:2015,Labro:2015,Bouji-etal:2020}. Since mantle radiogenic heat production controls how much heat is extracted from
 the core, it will also influence the presence or absence of a dynamo. Similarly, heat production will control
  the mantle temperature and thus the rate of silicate melting and volcanism. 
  
  Radiogenic heat production in the mantle is thus a key driver of rocky planet dynamics. \textcolor{black}{Dynamics in turn affects planetary habitability \citep[e.g.,][]{Lamme-etal:2009}; for instance, atmospheric evolution depends on both volcanism and dynamo activity.} As we argue below, rocky exoplanets may experience 
  large variations in radiogenic heat production, from $\sim$30\% to $\sim$300\% of the terrestrial value.
  In this paper we explore 
  the consequences of such variations for rocky planet dynamics and habitability .

\section{Radiogenic Element Origins and Distribution} \label{sec:elements}

The elements that make up the rocky planets were formed by \textcolor{black}{several} nucleosynthetic pathways operating in precursor stars. Following dispersal into the interstellar medium and -- ultimately -- into a molecular cloud, 
when the cloud collapses to form a planetary system, these elements are incorporated mostly into the new star but also into smaller 
solid bodies of progressively increasing size. Such planetesimals may preferentially lose volatile elements (those with low condensation temperatures,  {such as potassium}),  as the solid-body assembly process typically involves periods of high temperature. Nonetheless, one would expect the planets' final element concentrations
to resemble those of the star they orbit,  {especially for refractory lithophile elements like Th and U}. Thus, chemical abundance measurements of Galactic stars can be used to infer how the concentrations of various elements in their attendant planets are likely to vary, either spatially or in time. 

From the point of view of {long-term} radiogenic heating, the most important elements are potassium, thorium, and uranium.
Th and U are both r-process elements, most likely produced in
neutron star mergers {(NSMs)} \citep{Kasen-etal:2017,Cowan-etal:2019},  {with possibly significant contributions also from massive star collapse explosions} \citep{Siegel:2019,Macias:2019}. Because such events are rare (an occurrence rate of $\lesssim$ tens/Myr in our Galaxy) \textcolor{black}{and produce large quantities of r-process elements}, the resulting concentration of r-process elements should vary considerably. \textcolor{black}{This is especially true of low-metallicity stars,} \textcolor{black}{because turbulent mixing of the interstellar medium is inefficient in the early Galaxy} \citep{Shen-etal:2015,Naiman:2018}. This expectation is in agreement with 
measurements of r-process europium in low-metallicity stars \citep[e.g.,][]{2018ApJ...860...89M} and with r-process abundances in dwarf galaxies \citep[e.g.,][]{2016Natur.531..610J}. 

\textcolor{black}{Higher-metallicity stars also show larger variations of r-process elements than of elements (such as Mg) produced in core-collapse supernovae \citep[e.g.,][]{Cowan-etal:2019}}. Measurements of short-lived r-process nuclide decay products \citep{Hotok-etal:2015} and relative concentrations \citep{BartoMarka:2019} on Earth and in meteorites give further credence to the idea of a spatially {inhomogeneous} r-process distribution created  {by rare events such as NSMs.}

The r-process element europium, which is \textcolor{black}{easier to observe} in stellar spectra \textcolor{black}{than U or Th}, is commonly used to measure the r-process content in stars and, as expected, appears to be a robust proxy for thorium \citep{Botelho:2019} and uranium \citep{2008ARA&A..46..241S}, despite complications arising from decay of \textcolor{black}{$\rm ^{235}U$} \textcolor{black}{and severe blending of the Th~II line}.  The quantity [Eu/H] shows a range of -0.5 to +0.5 for Milky Way disk stars \textcolor{black}{ and shows a roughly linear correlation with [Fe/H]} \citep{Battistini:2016}. \textcolor{black}{ For thermal evolution models, what matters is the rate of heat production per unit mass of silicates. As a result, the most relevant metric is the concentration of Eu (a proxy for U and Th) compared with that of the alpha elements Mg and Si (proxies for the bulk silicate mantle).} The \textcolor{black}{10-90\% percentile} range observed in [Eu/Mg] 
is from about -0.3 to +0.3
for \textcolor{black}{likely planet-hosting stars} (i.e., those with [Mg/H] $>-0.3$)
\citep{Battistini:2016,DelgadoMena:2017,DelgadoMena:2019,Griffith:2019}, \textcolor{black}{while based on the Hypatia catalog (https://www.hypatiacatalog.com/), the 1-99\% range is about -0.5 to +0.5, or 30\% to 300\% \citep{Hinkel-etal:2014}}.  
The Eu/Mg ratio is likely to be a good indicator of the variation in \textcolor{black}{volumetric heat production} between planets orbiting different stars in the disk \textcolor{black}{because of the expected correspondence between host star and planet compositions.}


In contrast to Th and U, potassium is produced in type II supernovae ($\rm ^{39}K$; \cite{Shima-etal:2003}) and via s-process nucleosynthesis ($\rm ^{40}K$; \cite{The-etal:2007}). Because neither pathway is rare, spatial variations in potassium between stars are expected to be limited \citep{Lugaro:2018}. However, rocky bodies in our Solar System exhibit large variations in K
  concentrations \citep{McCub-etal:2012} because K is much more volatile, and thus more susceptible to loss during accretion, than U or Th. \textcolor{black}{Alpha elements such as Mg and Si which make up rocky planet mantles are neither volatile nor expected to show as much variability as r-process elements \citep{2008ARA&A..46..241S}. Stellar Mg/Si ratios show} only tens of percent variability 
  \citep{BrewerFischer:2018}, in agreement with measurements from polluted white dwarfs,\textcolor{black}{which provide estimates of planet composition} \citep{Doyle:2019}. 
 In summary, we expect Th and U to exhibit variations in concentration \textcolor{black}{relative to Mg} from 30 to 300\% of the Solar System value
 between different Sun-like stars in our Galaxy (and their associated planets). \textcolor{black}{While K might exhibit volatile-related variability between (or within) stellar systems, such variations are much harder to predict a priori, and the overall contribution of $\rm ^{40}$K to the present-day heat budget is in any case quite modest. Below we focus on} the effect on radiogenic heating of rocky planets with a wide range of Th and U concentrations.
  
 {The Earth's total present-day surface heat flow is in the range $\rm 42-47$~TW \citep{Sclat-etal:1980, Davies:2010}, or $\rm 35-40$~TW when the radiogenic contribution of the crust is subtracted  \citep{Sclat-etal:1980,O'Neill:2020}. The total radiogenic heat production of the Earth is about 22~TW, based on chondritic meteorites \citep[reviewed in][]{O'Neill:2020}, of which about 15~TW \textcolor{black}{is sourced within the mantle}.  These are similar to what we assume for our fiducial Earth model discussed below.  However, we note that geoneutrinos from the $^{238}$U and $^{232}$Th decay chains detected by Borexino have recently led to a much higher estimate of the Earth's total radiogenic heat of $38.2^{+13.6}_{-12.7}$ TW, and a total mantle heat contribution of $24.6^{+11.1}_{-10.4}$ TW \citep{Agostini:2019}. }

 \section{Thermal Evolution} \label{sec:thermal}

Below we carry out some simple explorations of how the initial radiogenic element concentration affects rocky planet evolution. 
  We focus on Earth-mass planets \textcolor{black}{with the same terrestrial mineralogy} and use a parameterized convection approach \textcolor{black}{modified from} \cite{Nimmo-etal:2004}, 
  in which plate-tectonic-like heat transfer is assumed. Because planetary dynamos are more likely to arise when plate tectonics is
   operating \citep{Nimmo:2002}, we do not address the contentious issue of whether plate tectonic or stagnant lid convection is more 
   likely on exo-Earths \citep[e.g.,][]{Tackl-etal:2013,NoackBreue:2014,WelleLenar:2016}. 

   Fully 3D, and even 2D, fluid dynamical models of Earth's thermal evolution are computationally expensive. As a result, parameterized convection
   models, which provide an averaged 1D description of mantle evolution, are frequently used, and are found to provide a reasonable
   match to 2D or 3D models \citep[e.g.,][]{VanKe:2001,Thiriet-etal}, \textcolor{black}{though they cannot capture the full complexity of a numerical model}. In the parameterized approach of \cite{Nimmo-etal:2004}, heat fluxes at both the top and bottom of the mantle are calculated by assuming that the local 
   (boundary layer) Rayleigh number is at the critical value for convection. \textcolor{black}{Based on the suggestion of \citet{Davies:2007} we modified the original bottom heat flux calculation (see Appendix); this produces qualitatively different results to the unmodified approach. Our modified model yields results that are qualitatively consistent with the numerical convection models of \citet{Plesa-etal:2015} for Mars, but should be regarded as somewhat provisional until they are checked by future numerical experiments.} 
   
   Mantle viscosity is temperature-dependent and is set to
    a reference value at a reference temperature; the reference viscosity at the bottom is higher than that at the top to mimic the effects 
    of increased pressure and phase transitions. The initial temperatures of the core and mantle are set to be the same at the core-mantle 
    boundary (CMB).  The temperature gradient in the mantle is assumed to be adiabatic, because individual convecting parcels do not have time to
    exchange heat with their surroundings. 
    
    The Earth's dynamo arises from fluid motions in its iron core. 
    These fluid motions are acted on by the Earth's rotation, but are ultimately
    driven by some combination of thermal and compositional buoyancy. The thermal buoyancy occurs because of extraction of heat from the core
    into the mantle; compositional buoyancy occurs as the solid inner core grows and releases light elements (e.g., Si or O) into the fluid above. Solid inner
    core growth starts only when the core has cooled sufficiently, so that for much \textcolor{black}{($\sim$60-90\%)} of Earth history a solid inner core was absent.
        As reviewed by \citet{Nimmo:2015,Labro:2015}, maintaining
    a well-mixed, convecting core in the absence of compositional buoyancy requires that the heat extracted from the core must exceed the adiabatic value. A more general
    approach (which includes the effect of composition) is to track the rate of entropy production rather than heat extraction \citep[e.g.,][]{Nimmo-etal:2004}.     Tracking the net entropy production rate $\Delta \dot{E}$ thus provides a simple proxy for whether a dynamo can operate or not, while the dipole field strength is expected to scale as $\Delta \dot{E}$ to the one-third power \citep{Labro:2015}.
    
    Figure 1b (middle panel) illustrates our baseline result, with parameters chosen so that the present-day known
    characteristics of the Earth (mantle temperature and viscosity structure, heat flux budget, inner core radius)  and petrological
   constraints on mantle temperature evolution \citep{Abbot-etal:1994} are reproduced. Although this particular set of parameter choices
    (notably the thermal conductivity of the core) is non-unique, the main point of this exercise is to isolate the effect of changing one particular parameter (radiogenic heat production) 
     and to examine its consequences.

\begin{figure}[h!tb]
\centering
\includegraphics[width=7.in]{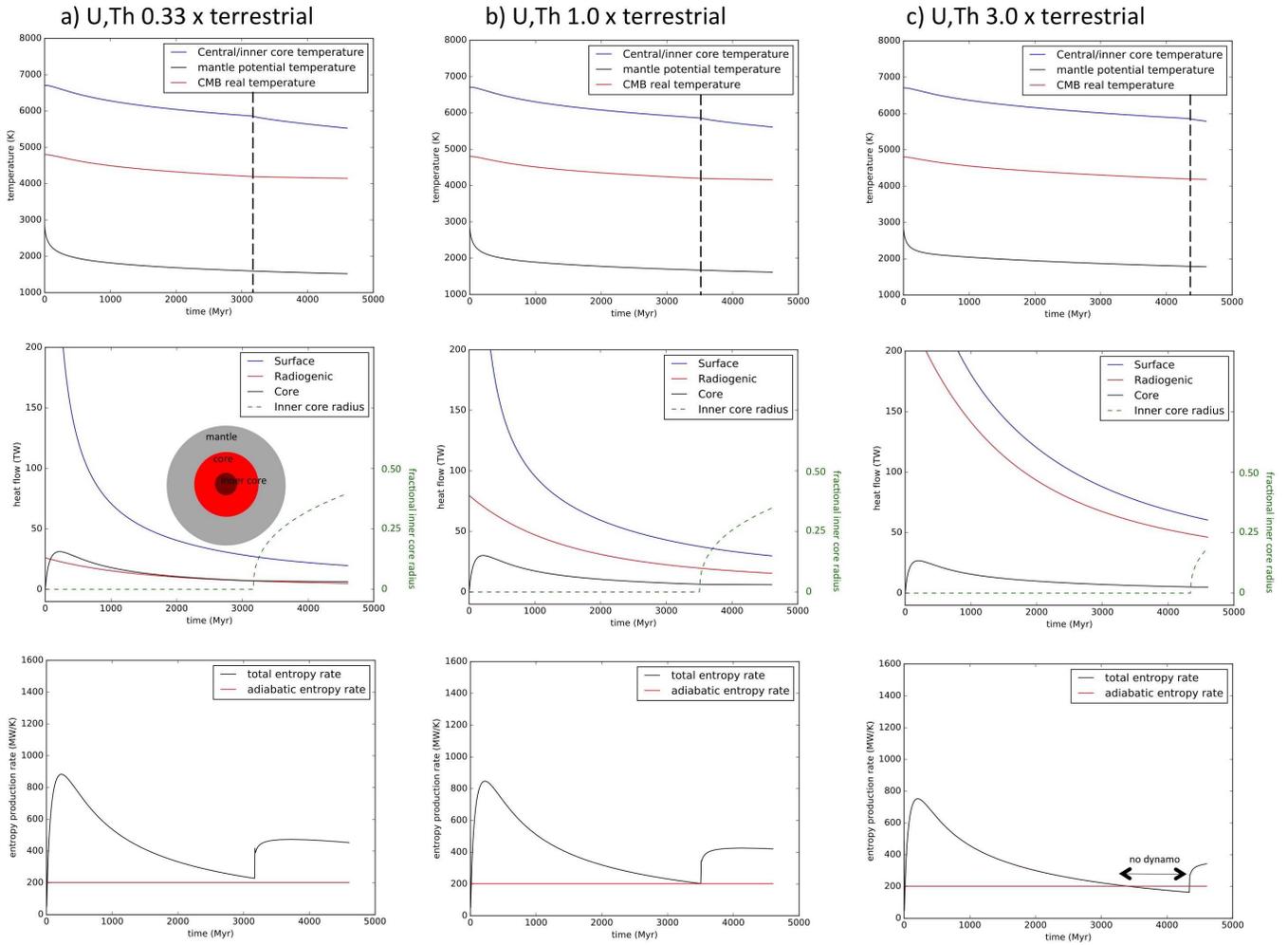}
\caption{Effect on thermal evolution and dynamo for different concentrations of radiogenic elements (U,Th): a) 33\% of terrestrial;
b) 100\% of terrestrial; c) 300\% of terrestrial. In each panel the top plot shows temperature evolution, the middle plot shows heat flow
and solid inner core radius, and the bottom plot the net rate of entropy production (driving a dynamo requires a 
value greater than the adiabat). Present-day
is at the end of the plots. The inset in panel a) shows the internal structure of the Earth. Vertical lines in top panels denote onset of inner core solidification.}
\end{figure}

The top panel shows the decline in core and mantle temperatures as a function of time, with the
      present-day mantle temperature projected to the surface (the potential temperature) $T_p$ of \textcolor{black}{1341}$\rm ^\circ$C and cooling rate of \textcolor{black}{50}~K/Gyr in 
      approximate agreement with petrological constraints \citep{Abbot-etal:1994}. The middle panel shows heat flow evolution, with a present day
       mantle radiogenic heat production rate of 15~TW and  \textcolor{black}{surface and core heat flows of 30 and 6}~TW, respectively. It also shows the growth of the inner core, which starts
       to solidify at 3515~Myr (1.05)~Ga as the core cools. The lowermost panel shows that the resulting rate of entropy 
       production exceeds the adiabatic value and is thus sufficient to maintain a dynamo, except possibly for a brief period just prior to inner 
       core formation. The entropy production rate increases markedly once the inner core starts to grow, because compositional buoyancy is now
       aiding thermal buoyancy. Paleomagnetic measurements show that a terrestrial dynamo has operated over at least the last 3.5 Gyr \citep{Tardu-etal:2010}, 
       \textcolor{black}{with at least one interval lasting} up to a few~Myr of low- or no-field \citep{Bono-etal:2019}.
       
       Because of the energetic nature of Earth accretion, the appropriate initial conditions are highly uncertain. Fortunately, they turn out not 
       to be very important: because mantle heat transport is highly dependent on temperature, mantle temperature evolution is controlled 
       mainly by radiogenic heat production. Thus, increasing the start temperature by 500~K changes the present-day potential temperature 
       by less than \textcolor{black}{1.5}~K; \textcolor{black}{\citet{Schubert-etal:1980} found an almost identical result.} Starting with a core hotter than the mantle likewise has little effect at the present day.  

Figures 1a and 1c show how an identical model evolves with the same initial conditions except for 0.33 and 3 times the assumed terrestrial
        concentrations of U and Th. Less and more radiogenic heating result in colder and warmer present-day mantles ($\Delta T_p$) of \textcolor{black}{-99}~K and \textcolor{black}{+166}~K),
         respectively; these results resemble those of \citet{Unter-etal:2015}. \textcolor{black}{The behavior of the core is more complicated, because of the competing effects of changes in the core-mantle temperature contrast and  mantle viscosity (see Appendix). The former effect alone would decrease the core heat flux in a hotter mantle, while the latter would increase it.  Somewhat counter-intuitively, we find that more radiogenic heating results in a lower present-day core heat flux, a smaller inner core and a less vigorous dynamo, and vice versa.    The case with enhanced mantle heating (Fig.~1c) experiences a dynamo that ceases to operate for hundreds of Myr.}
           
 \begin{figure}[h!tb]
\centering
\includegraphics[width=7.in]{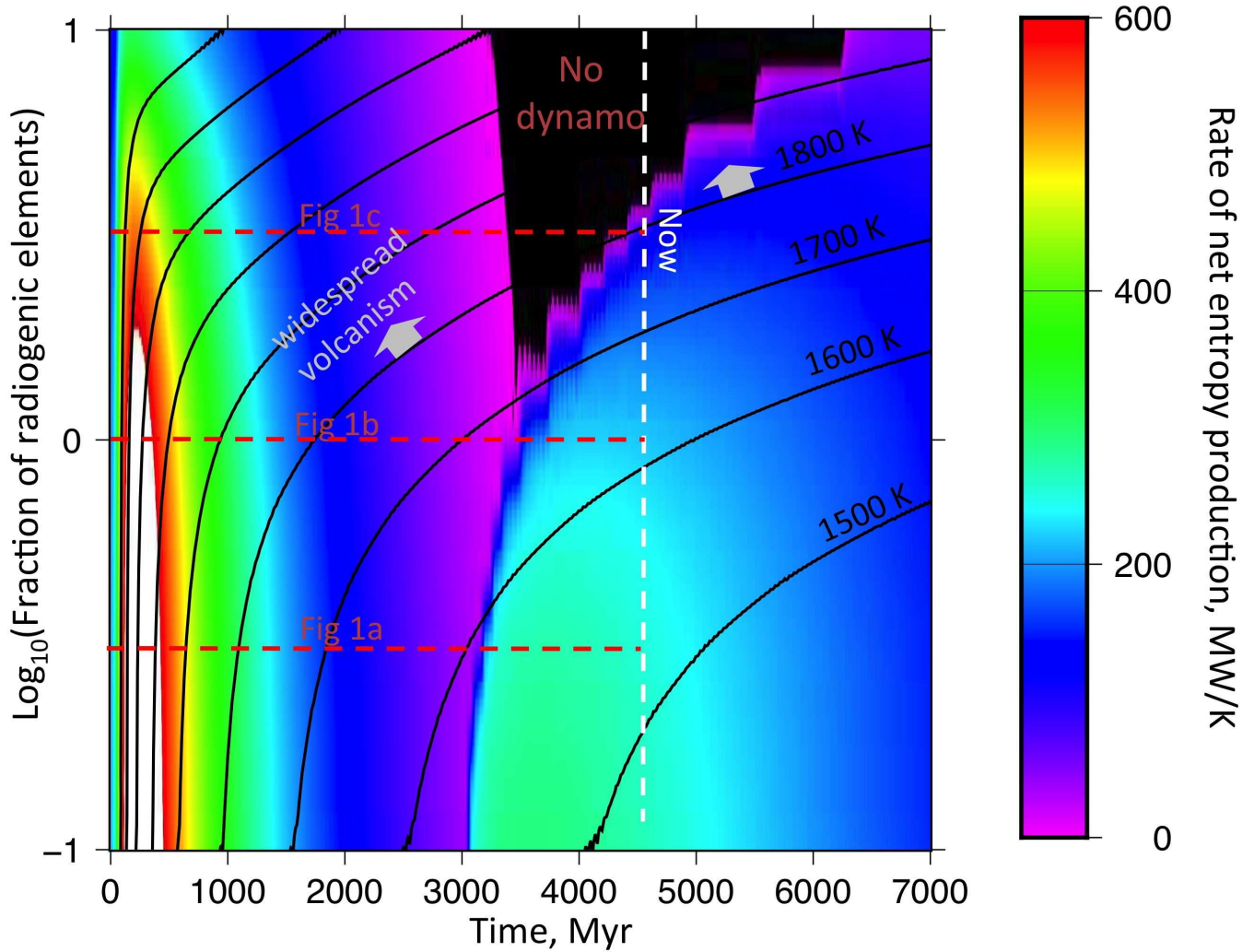}
\caption{Sensitivity of evolution of core parameters to different radiogenic element concentrations (relative to the nominal terrestrial case). The colors show the rate of net entropy production, with black indicating a negative value (no dynamo). \textcolor{black}{The contours denote the mantle potential temperature}. The three dashed red lines show the trajectories of the three evolution scenarios shown in Fig~1. Vertical white dashed line indicates present day.}
\end{figure}

           Figure~2 shows the net rate
           of entropy production (colors: a proxy for dynamo strength) as a function of time and radiogenic element fraction. The three trajectories shown
           in Figure~1 are marked as horizontal dashed lines, and the contour lines denote the \textcolor{black}{mantle potential temperature. As noted above,
           higher radiogenic element concentrations result in hotter mantles, smaller present-day inner cores and longer periods in which a dynamo is absent. }
           
\newpage

\section{Summary} \label{sec:Summary}

Our simplified model shows that higher concentrations of U,Th have two principal consequences: one is hotter present-day mantles; 
           the other is \textcolor{black}{reduced} dynamo activity. The converse is true for lower U,Th concentrations. Both of these effects are likely to 
           have major implications for habitability.

A global magnetic field modifies the trajectories of charged particles emitted by the host star. In our Solar System the net effect of such 
           a field is probably to reduce rates of long-term atmospheric loss due to particle bombardment \citep{Lundi-etal:2007} though this is 
           debated \citep{Gunel-etal:2018}. How such effects would translate to close-in Earth-like exoplanets is uncertain \citep{Owen:2019} but,
           overall, dynamo activity is regarded as an important component of planetary habitability \citep{Lamme-etal:2009}.
           
           Assuming a similar bulk mantle composition (in terms of major elements: Mg, Si etc.), higher mantle temperatures will yield more melting \textcolor{black}{(and thus volcanic activity)} and higher convective velocities, both of
which will produce higher melt generation rates. Because volatile elements (like water)
            tend to concentrate in silicate melts, higher total melt production will result in more efficient outgassing from the interior.  The likely 
            consequence is a more volatile-rich surface and/or a thicker atmosphere, depending on the degree of volatile return to the interior 
            by plate tectonics \citep[e.g.,][]{NoackBreue:2014,FoleySmye:2018}.

           Early discussions of planetary habitability focused on the orbital semi-major axis \textcolor{black}{and stellar flux} \citep[]{Hart:1979}. More recent investigations have added other factors of interest \citep[e.g.,][]{Lamme-etal:2009}, \textcolor{black}{including the abundance of r-process elements \citep[e.g.,][]{Frank-etal:2014} which we argue here will help determine both dynamo and volcanic outgassing activity}. An important aspect of this particular factor is that {\em it can be determined by observations}. As discussed above, planetary concentrations of U and Th are expected to track those of the host star, which can be measured via spectroscopy \textcolor{black}{through the surrogate element Eu} \citep[e.g.,][]{Botelho:2019,GALAH:2020}.
                 For instance, 
               \citet{daSilva:2016}
              shows an increase in Eu (and by inference Th and U) relative to Fe of roughly 0.5~dex from the Galactic center to the outer disk. We conclude that stellar abundances of K, Th, and U 
              should be taken into account in designing future observational campaigns.
               

\section{Discussion/Future Work} \label{sec:Discussion}

       The calculations shown in Fig~1 and 2 are  highly simplified. In particular, although they are tuned to match existing constraints on Earth's present-day state and evolution, they become progressively more uncertain at earlier times, when the initial conditions chosen are more important and unmodeled processes (e.g., melt advection, different tectonic styles) may operate. \textcolor{black}{Furthermore, as noted above, even the qualitative behavior of the core heat flux is sensitive to  particular modeling choices (see Appendix). 2D or 3D convection models investigating the role of radiogenic heat production on exoplanet heat transfer -- along the lines of \citet{ONeill-etal:2020} -- can capture details (e.g., the effect of phase transitions) not treated here, and resolve current uncertainties in the link between radiogenic heating and core heat flux.}

             
             
             The models shown in Fig.~1 all assume plate-tectonic like heat transfer. But the example of Venus shows that even for Earth-like planets, plate 
             tectonics is not assured. Because plate tectonics cools the mantle (and thus the core) efficiently, long-term plate tectonics is likely 
             crucial to the maintenance of a dynamo \citep{Nimmo:2002}. Unfortunately, exactly what determines whether plate tectonics occurs is not
              understood; the presence of water (and thus the eccentricity and semi-major axis of the planet's orbit) and/or melt are likely important \citep{Loure-etal:2018}, \textcolor{black}{and so too are the mantle driving stresses, and thus the heat production rate \citep{ONeill-etal:2020}}. Since the survival of water and the generation of melt are in turn both sensitive to the radiogenic abundance,
              predicting how these factors interact with each other is not straightforward.  

             How would planetary radius influence our conclusions? A larger planet produces more radiogenic heat per unit area; to first order the heat fluxes will therefore scale with radius $R$. But the minimum (adiabatic) heat
             flux required to drive a dynamo also scales with gravity, and thus also with $R$. So larger planets are not guaranteed to have more vigorous dynamos \textcolor{black}{(though we do expect them to be more volcanically active)}; more detailed treatments are required, such as those of \citet{Bouji-etal:2020} but including radiogenic heating.
             
             Spatial and temporal variability in radiogenic element concentrations will also play a role. As noted above, U and Th concentrations are expected to increase towards the edge of the Galactic disk 
             \citep{daSilva:2016}. The age of the planet is more complicated. Obviously, other things being equal an older planet will have lower radiogenic heat production (because the nuclides will have decayed) and a weaker or absent dynamo. Furthermore, \citet{DelgadoMena:2019} show that the stellar Eu/Mg ratio decreases with increasing stellar age \textcolor{black}{at about 0.1 dex per 10~Gyr}, compounding this effect. \textcolor{black}{In \citet{ONeill-etal:2020} the stellar age is assumed to cause the core size (via the ratio Mg/Fe) and radiogenic heat production to co-vary. By contrast, in our view} the impact of any secular variation in initial U,Th concentrations will be less important than the expected random scatter.  \textcolor{black}{In any event, more general explorations of the role of r-process elements in determining planetary behavior, and the specific influence of core size, planet size and plate tectonics on habitability, are of interest to pursue in future.}

             \textcolor{black}{Our results are summarized in Fig.~2: higher mantle heating rates increase volcanic activity and shut off the dynamo, both likely deleterious for habitability. In contrast, reduced mantle heating rates will at some point stop melt production. A potential consequence is that plate tectonics will then cease \citep{Loure-etal:2018}, and with it, the dynamo, as at Venus \citep{Nimmo:2002}.} It is tempting to speculate that the Earth is habitable in part because it possesses a ``Goldilocks'' concentration of radiogenic elements: high enough to permit long-lived dynamo activity \textcolor{black}{and plate tectonics}, but not so much that extreme volcanism and dynamo shutoff occur. As yet, however, our understanding of the complicated feedbacks involved is too rudimentary to make such a definitive conclusion; more detailed calculations, and better characterization of the radiogenic element abundances in planet-hosting stars, are both required.


 \vspace{1.cm}
              
   Acknowledgments: E.R.-R.  thanks  the  Heising-Simons  Foundation,  and  the  Danish  National  Research  Foundation (DNRF132)  for  support. SMF acknowledges support from NSF AST-1615730.   We thank Natalie Batalha, Ben Zuckerman, \textcolor{black}{Matt Shetrone, David Weinberg, Ina Plesa, Steve Ritz and Charli Sakari} for helpful discussions, \textcolor{black}{and the reviewer for insightful comments.}      
   


\appendix


Below we provide some additional details on the implementation of the parameterized convection;
parameter values not discussed here are the same as in \citet{Nimmo-etal:2004}. We also note here that in our analysis of the Hypatia
catalog \citep{Hinkel-etal:2014} we exclude the data of { Luck et al.} (2015,2017,2018) from our results, since these show anomalously high Eu concentrations. Thus, the range for [Eu/Mg] of -0.5 to +0.5~dex is conservative and would be higher if the Luck data were included.

Following \citet{O'Neill:2020} we assume initial Bulk Silicate Earth concentrations of
260~ppm, 23~ppb and 85~ppb respectively for K, U and Th; this produces 22~TW of heat
production at the present day. \textcolor{black}{We implicitly assume that the major element chemistry and mineralogy is the same as that of Earth; these variables are expressed in the model via the reference viscosities and activation energy assumed for the mantle.} We assume that the convecting mantle is responsible for 70\% of
the total radiogenic heat at all times; the remainder is assumed to reside in the crust and will not contribute to mantle thermal evolution. In examining
the effect of varying radiogenic element concentrations, we allow U and Th concentrations
to vary from their nominal value but keep the K concentration the same, because only r-process
elements are expected to show large variations from star to star (see main text).

As discussed in more detail below, heat transfer across the top and bottom boundary layers of the mantle is calculated by
using a local Rayleigh number approach. \textcolor{black}{The heat flux out of the core is affected by two competing processes. A hotter mantle has a lower viscosity, and thus a tendency to increase the heat flux; but it also decreases the core-mantle temperature difference, lowering the heat flux. Because of these competing effects, we find that small changes in model assumptions can lead to qualitatively different behavior. We adopt an approach suggested by \citet{Davies:2007} (see below), but note that future numerical models are required to check this approach. }

The \textcolor{black}{bottom} boundary layer thickness $\delta_b$ is given by
\[
\delta_b = \left[\frac{Ra_c \kappa_b \eta_b}{\rho_m g \alpha_m (T_c-T_m)} \right]^{1/3}
\]
where $Ra_c$ is the critical Rayleigh number (set to 600), $\kappa_b$, $\rho_m$ and $\alpha_m$
are the mantle thermal diffusivity, density and thermal expansivity, $g$ is the acceleration due to
gravity and $T_c$ and $T_m$ are the core temperature at the core-mantle boundary (CMB) and the mantle temperature projected down to the CMB along the adiabat, respectively.

\textcolor{black}{The viscosity $\eta_b$ was evaluated by \citet{Nimmo-etal:2004} at the mean boundary layer temperature ($T_a=[T_c+T_m]/2$). But it was argued in \citet{Davies:2007} that for a bottom boundary layer it is more appropriate to evaluate the viscosity at $T_c$ instead, and we follow this approach here.} 

The basal viscosity at a temperature $T$
is evaluated according to
\[
\eta_b(T) = f \eta_0 \exp[-\xi (T-T_0)]
\]
where $\eta_0$ is the reference viscosity at a reference temperature $T_0$, \textcolor{black}{$f$ is a factor used to account for the intrinsically higher viscosity of the lower mantle,} and $\xi$ describes the
temperature-sensitivity of the viscosity and is taken to be $\rm 0.01~K^{-1}$ here. At the nominal present-day reference viscosity and mantle temperature of \textcolor{black}{$\rm 1341^\circ$~C} the upper and lower mantle viscosities
are $\rm \textcolor{black}{1.3}\times10^{21}$ and $\rm \textcolor{black}{1.6}\times10^{22}$~Pa~s.

\textcolor{black}{The CMB heat flux is then given by $F_b = k_b (T_c - T_m)/\delta_b$ where $k_b$ is the thermal
conductivity at the base of the mantle. Because of the sensitivity of the core heat flux to two opposing effects (see above), heat flux results using $\eta_b(T_a)$ yield quite different results to those using $\eta_b(T_c)$. Uncertainties in what reference viscosity $\eta_0$ to assume, while large, are less important than deciding whether to use $\eta_b(T_c)$ or $\eta_b(T_a)$ because we can determine the reference viscosity by needing to reproduce the present-day Earth characteristics. }

\textcolor{black}{We adopt the $\eta_b(T_c)$ approach advocated by \citet{Davies:2007} because it produces results which are qualitatively consistent with the numerical models of \citet{Plesa-etal:2015}. Specifically, these authors find that higher mantle heat production rates yield higher mantle and core temperatures and (in general) lower time-averaged core heat fluxes, which agrees with our findings. Furthermore, our heat flux results are in line with the numerical models of \citet{ONeill-etal:2020}, though the latter may still be experiencing initial transients.}
    
The surface heat flux is calculated using similar expressions except that: the driving temperature difference
is $T'_m-T_s$ where $T'_m$ and $T_s$ are the mantle potential temperature and surface temperature; and the thermal
diffusivity and thermal conductivity are assumed lower than in the lower mantle.

The core thermal conductivity is here taken to be $\rm 50~W~m^{-1}~K^{-1}$ and 100~ppm potassium
is assumed to be present in the core. Parameter values different from the nominal values used in \citet{Nimmo-etal:2004}
are as follows: mantle bulk density $\rm \rho_m = 4400~kg~m^{-3}$, reference viscosity $\rm \eta_0 = 2\times10^{21}~Pa~s$, 
lower mantle viscosity factor $f$=20, \textcolor{black}{and the reference temperature for the lower mantle is $T_0=4070$~K.}


\bibliography{refs}{}
\bibliographystyle{aasjournal}



\end{document}